\documentclass[showpacs,showkeys,twocolumn]{revtex4}

\usepackage{array}
\usepackage{amsmath,amsfonts,amssymb,dsfont,mathrsfs,centernot,amsthm}
\usepackage{graphicx}
\usepackage{comment}

\newcommand{\beq}{\begin{equation}}
\newcommand{\eeq}{\end{equation}}
\newcommand{\ber}{\begin{eqnarray}}
\newcommand{\eer}{\end{eqnarray}}

\newcommand{\df}{{\bf{d}}}

\newcommand{\J}{\mathscr{J}}

\newcommand{\Z}{\mathbb{Z}}

\newcommand{\ga}{\gamma}
\newcommand{\fy}{\centernot}

\newcommand\VIN[2]{E^{\hat{#1}}_{\hat{#2}}}

\newcommand\GAM[1]{\gamma^{\hat{#1}}}

\newcommand\NAB[1]{\nabla_{\hat{#1}}}

\newcommand\PA[1]{\partial_{\hat{#1}}}

\theoremstyle{remark}
\newtheorem*{rem}{Remark}

\begin{document}

\title{On the localization of fermions in double thick $D$-branes}
\author{Oscar Castillo-Felisola}
\email[E-mail address: ]{o.castillo.felisola@gmail.com}
\author{Iv\'an Schmidt}
\email[E-mail address: ]{ivan.schmidt@usm.cl}
\affiliation{Departamento de F\'\i sica y Centro Cient\'\i fico
Tecnol\'ogico de Valpara\'\i so, Universidad T\'ecnica Federico
Santa Mar\'\i a, Casilla 110-V, Valpara\'\i so, Chile.}

\begin{abstract}
  Hints on the possible localization of fermions on double thick $D$-branes (Domain Walls) are found by analyzing the moduli space of parameters. Deeper analysis toward this direction might help to select phenomenologically plausible models. A new kind of condition for fermion localization is proposed. This might be useful in multi-brane-world scenarios, which are important when symmetry breaking is considered in the AdS/CFT formalism, as well as  in curved brane-worlds. 
\end{abstract}


\maketitle



\section{Introduction}

In the last decade, after the work of Randall and Sundrum\cite{RS1,RS2}, the study of brane-worlds have increased considerably, due to its application in explaining fundamentals problems in both particle physics and gravity.

Besides the early works \cite{Akama:1982jy,RS,Visser:1985qm}, the importance of brane-worlds became apparent almost 20-years later, when the hierarchy problem was considered in this context \cite{RS1}, and the same authors realized that a new kind of compactification was possible \cite{RS2}, different to that proposed by Kaluza and Klein \cite{Kaluza,Klein} and its generalizations \cite{PopeKK}.

On the other hand, branes are the fundamental brick for the AdS/CFT correspondence, proposed by Maldacena\cite{Maldacena1} (see also \cite{Maldacena2}), which allows to relate the spectrum of a string theory in a AdS background with the one of a conformal field theory at the boundary of the AdS spacetime. Throughout this article only the brane-world aspect will be considered.

Several generalizations of the Randall-Sundrum brane worlds have been considered, such as multi-brane worlds\cite{Kogan:2000xc}, thick branes or domain walls\cite{Gremm:1999pj}, more complicated domain walls solutions with or without a well defined brane-limit\cite{MPS,Guerrero:2005aw,Guerrero:2005xx} (for an illustration on the brane limit see \cite{GMP}), localization of gravity in different brane-world backgrounds\cite{CastilloFelisola:2004eg}, localization of fermions\cite{Kehagias:2000au,MPT} and their spectrum\cite{Liu:2009dw}, localization of gauge fields\cite{Guerrero:2009ac} and many others.

The aim of this paper is to restrict the moduli space to specific variety of domain walls, known in the literature as {\it static double walls} \footnote{This kind of solutions is known as double domain wall because are not just  kink-like solitons but look like a stack of two kinks. Moreover, the Kretschmann scalar has a couple of singular hyper-surfaces in the neighborhood of the extra dimension.}, found in \cite{MPS} and generalized in \cite{Guerrero:2005aw,Guerrero:2005xx}. Despite the existence of these generalizations, we concentrate ourselves on the original solutions.  The motivation for studying these static domain walls is that they are the minimal parametric generalization of the thick version of Randall-Sundrum brane-world \cite{Gremm:1999pj}, which are suitable  phenomenological candidates for allowing  fermion localization.

In the following section a detailed analysis of the localization of fermions is done, for static solutions. Although the procedure is known and reviewed in several articles, new considerations are made and special cases considered. In section \ref{sec:static} the analysis of the spectrum and moduli space is done for a family of static double domain walls. At the end an appendix is included to summarize notation and conventions.

\nocite{sage}

\section{Fermion Localization on Static $D$-branes}\label{sec:loc}

In the following, thick-branes (or gravitational domain walls) are considered. The formal relation between the two might be established through a limit \cite{GMP}.

This Domain Walls are generated by a scalar field, $\phi(\xi)$, which depends on the extra dimension, $\xi$ (see \cite{Vilenkin-book}). The spacetime has a metric whose line element is given by \footnote{Since only static solutions are considered in the following, the $B$ function  is considered as $t$-independent.},
\begin{equation}
ds^2(g)=e^{2A(\xi)}\left(-dt^2+e^{2B(\xi,t)}d\vec{x}^2\right)+e^{2C(\xi)}d\xi^2.
\end{equation}
Several Domain Wall solutions are listed on \cite{MPS,Guerrero:2005aw,Guerrero:2005xx}.

The 5-dimensional action for a massive fermion on a curved spacetime is given by,
\begin{equation}
  S_D = \int d^4x d\xi \sqrt{|g|}\bar{\Psi}\left(-\not\!\nabla - M
\right)\Psi,
\end{equation}
where $\Psi$ is the 5-dimensional fermion,  the bar indicates Dirac conjugation, $\nabla$ is the covariant derivative on the curved spacetime, $M$ is the 5-dimensional mass of $\Psi$ and $|g|$ is the absolute value of the metric determinant. The use of tetrad formalism is implicit, although the notation does not suggest it.


However, it is not possible to localize fermions  on this set up \cite{Bajc:1999mh}. A bypass to the problem consist in adding an interaction. A Yukawa-like interaction between the fermion an the scalar field which generates the domain wall does the work. Henceforth, the action to consider is,
  \begin{equation}
    S = S_D + S_Y = \int d^4x\;d\xi \sqrt{|g|}\;\bar{\Psi}\left(-\fy\nabla-M +  \lambda\mathcal{P}(\phi)\right)\Psi,\label{Dirac-act}
  \end{equation}
with $\mathcal{P}$ is, in general,  a polynomial function on $\phi$. The simplest model is $\mathcal{P}(\phi) = \phi$, and it was considered in \cite{Bajc:1999mh} with a different kind of Domain Wall than those presented here.


In the literature, it has become customary to take into account just the case when the 5-dimensional mass of the fermion vanishes, i.e. $M=0$. However, as will be discussed, $M$ could be a rather important parameter for selecting plausible phenomenological models. Therefore,  the distinction between massive and massless fermions is made. 

\subsection{Massless 5D Fermions}

In the following section a presentation of the different possible conditions which could be imposed on the fermions are given, as well as the equations of motion of their profiles \footnote{Some of these choices might be too strong in general spacetimes. Nonetheless, under the right circumstances (whose study is beyond the scope of this report), the problem could be well-posed.}.

The static Domain Wall solution given in \cite{MPS} has a
$B(\xi)\propto \xi$, which vanishes at the wall position. This
favors  to impose the `flat' condition on 4-dimensional fermions, as
explained on sections \ref{sec:flat1} and \ref{sec:flat2} below.
However, in multi-brane scenarios \cite{RS1} or in curved branes \cite{Papadopoulos:1999tw}, a different
4-dimensional condition could be imposed, as mentioned in sections
\ref{sec:cov1} and \ref{sec:cov2} below.

From now on, the 5-dimensional spinor is decomposed into,
\begin{equation}
  \Psi(x,\xi) = \psi_+(x) f_+(\xi) +\psi_-(x) f_-(\xi),
\end{equation}
with $\psi_\pm= P_\pm\psi$ the 4-dimensional chiral spinors  and $f_\pm$ their profile through the extra dimension.

\subsubsection{Flat Massless 4D Fermions}\label{sec:flat1}

For a flat brane world volume, the Dirac equations for  4D chiral massless fermions, $\psi_\pm$, are 
\begin{equation}
  \not\! \partial^{(4)}\psi_\pm=0. \label{Dirac-chiral-massless}
\end{equation}
Following the procedure described in \cite{Kehagias:2000au,MPT}, one gets the equations of motion for the profiles,
\begin{equation}
f'_\pm + A'f_\pm \mp \lambda\mathcal{P}(\phi)e^C f_\pm =0,
\end{equation}
where $f_\pm$ are the profiles for $\psi_\pm$. Their solutions are,
\begin{equation}
  f_\pm \propto e^{-A\pm \lambda\int d\xi\;\mathcal{P}(\phi)e^C}.\label{prof-flat-massless4d}
\end{equation}

\subsubsection{Flat Massive 4D Fermions}\label{sec:flat2}

If the world volume of the brane is flat, but the 4-dimensional fermions are massive, the Dirac equation for these are,
\begin{equation}
  \not\!
  \partial^{(4)}\psi_\pm=-m\psi_\mp,
\end{equation}
with $m$ the 4-dimensional mass. A similar procedure to the one above,  yields
the profile equations  (see \cite{MPT}),
\begin{equation}
f'_\pm + A'f_\pm \mp \lambda\mathcal{P}(\phi)e^C f_\pm =\pm m e^{-A+C} f_\mp,
\end{equation}
since massive fermions mix chiralities, the equations are coupled.

The best way of work them out is changing the variable from
$\xi\to\xi'$ where $\xi'=\int d\xi e^{-A+C}$. Then, in terms of the
new variable one gets
\begin{equation}
f'_\pm + A' f_\pm \mp \lambda\mathcal{P}(\phi) e^A f_\pm = \pm m f_\mp,
\end{equation}
with prime denoting derivative with respect to $\xi'$.

When the equations are decoupled, a Schr\"odinger-like equation governs the profiles \footnote{As shown in \cite{MPT}.},
\begin{equation}
\left[-\partial^2_{\xi'} + V^\pm_{qm}\right]u_\pm = m^2 u_\pm,
\end{equation}
where $$f_\pm \mapsto e^{-2A}u_\pm,$$ and
\begin{equation}
V^\pm_{qm} = \left(\lambda \mathcal{P}(\phi)e^A\right)^2 \pm \partial_{\xi'}\left(\lambda \mathcal{P}(\phi)e^A\right).\label{flat-Vqm}
\end{equation}

\subsubsection{Covariantly Massless 4D Fermions}\label{sec:cov1}

There exist branes which are not flat submanifolds embedded on the spacetime \cite{Papadopoulos:1999tw}. For these general cases, the Dirac equation for 4-dimensional massless chiral fermions is given by
\begin{equation}
  \not\! \nabla^{(4)}\psi_\pm=0,
\end{equation}
and the profile equations are,
\begin{equation}
f'_\pm + \frac{1}{4}\left(4A'+{3}B'\right)f_\pm \mp \lambda\mathcal{P}(\phi)e^C f_\pm =0,
\end{equation}
whose solutions are,
\begin{equation}
  f_\pm \propto e^{-\frac{1}{4}\left(4A+3B\right)\pm \lambda\int d\xi\;\mathcal{P}(\phi)e^C}.\label{prof-cov-massless4d}
\end{equation}

\subsubsection{Covariantly Massive 4D Fermions}\label{sec:cov2}

On these curved branes, 4-dimensional massive fermions might be considered. Their Dirac equations are, 
\begin{equation}
  \not\! \nabla^{(4)}\psi_\pm=-m\psi_\mp,
\end{equation}
whilst the profile equations are,
\begin{equation}
f'_\pm + \frac{1}{4}\left(4A'+{3}B'\right)f_\pm \mp \lambda\mathcal{P}(\phi)e^C f_\pm =\pm m e^C f_\mp.
\end{equation}

Proceeding as before, but with the changes $\xi'=\int d\xi e^{C}$
and $f_\pm \mapsto e^{-\frac{4A+3B}{2}}u_\pm$, the
Schr\"odinger-like equation governing  the profiles is,
\begin{equation}
\left[-\partial^2_{\xi'} + V^\pm_{qm}\right]u_\pm = m^2 u_\pm,
\end{equation}
with
\begin{equation}
 V^\pm_{qm} = \left(\lambda \mathcal{P}(\phi)\right)^2 \pm \partial_{\xi'}\left(\lambda \mathcal{P}(\phi)\right).\label{cov-Vqm}
\end{equation}

\begin{rem}
  Equation (\ref{cov-Vqm}) has the same structure as (\ref{flat-Vqm}), but the lack of  the exponential functions make it looks simpler. However, non-flat domain-wall solutions are much more complicated than flat ones.
\end{rem}

\subsection{Massive 5D Fermions}

Although in theories with extra dimensions it is preferred to begin with massless fermions, there exists no fundamental reason for doing so. Thus, in case  5-dimensional massive fermions are considered, the above solutions  are valid after the substitution $\lambda\mathcal{P}(\phi) \to
\lambda\mathcal{P}(\phi)-M$, where $M$ in the 5-dimensional mass.

From (\ref{prof-flat-massless4d}) and (\ref{prof-cov-massless4d}),
it is clear that after the substitution $\lambda\mathcal{P}(\phi)\to
\lambda\mathcal{P}(\phi)-M$, the localization of the positive
profile, $f_+$, is favored in detriment of the negative one, $f_-$.
It is possible that a critical region exists where both chiralities
might be localized.

In the case of 4-dimensional massive fermions, the effect enters through the quantum mechanical potential, (\ref{flat-Vqm}) and (\ref{cov-Vqm}), which is a case less intuitive to analyze. However, these equations might be solved numerically. From their solutions one can argue whether or not the model is phenomenologically relevant. This analysis in done for the double domain walls in the next section.

\section{Double Domain Wall}\label{sec:static}

In the special case of the so called double walls \cite{MPS}, whose metric has a line element given by
\begin{equation}
  ds^2(g)=e^{2A(\xi)}\left(\eta_{\mu\nu}d x^\mu d x^\nu+d\xi^2\right),
\end{equation}
with a warp factor,
\begin{equation}
  A(\xi)=-\frac{1}{2s}\ln \left(1+(\alpha \xi)^{2s}\right),
\end{equation}
the scalar field is,
\begin{equation}
  \phi(\xi)=\frac{\sqrt{6s-3}}{s}\arctan\left(\alpha\xi\right)^s.
\end{equation}

\begin{figure}
   \begin{center}
       \includegraphics[scale=0.33]{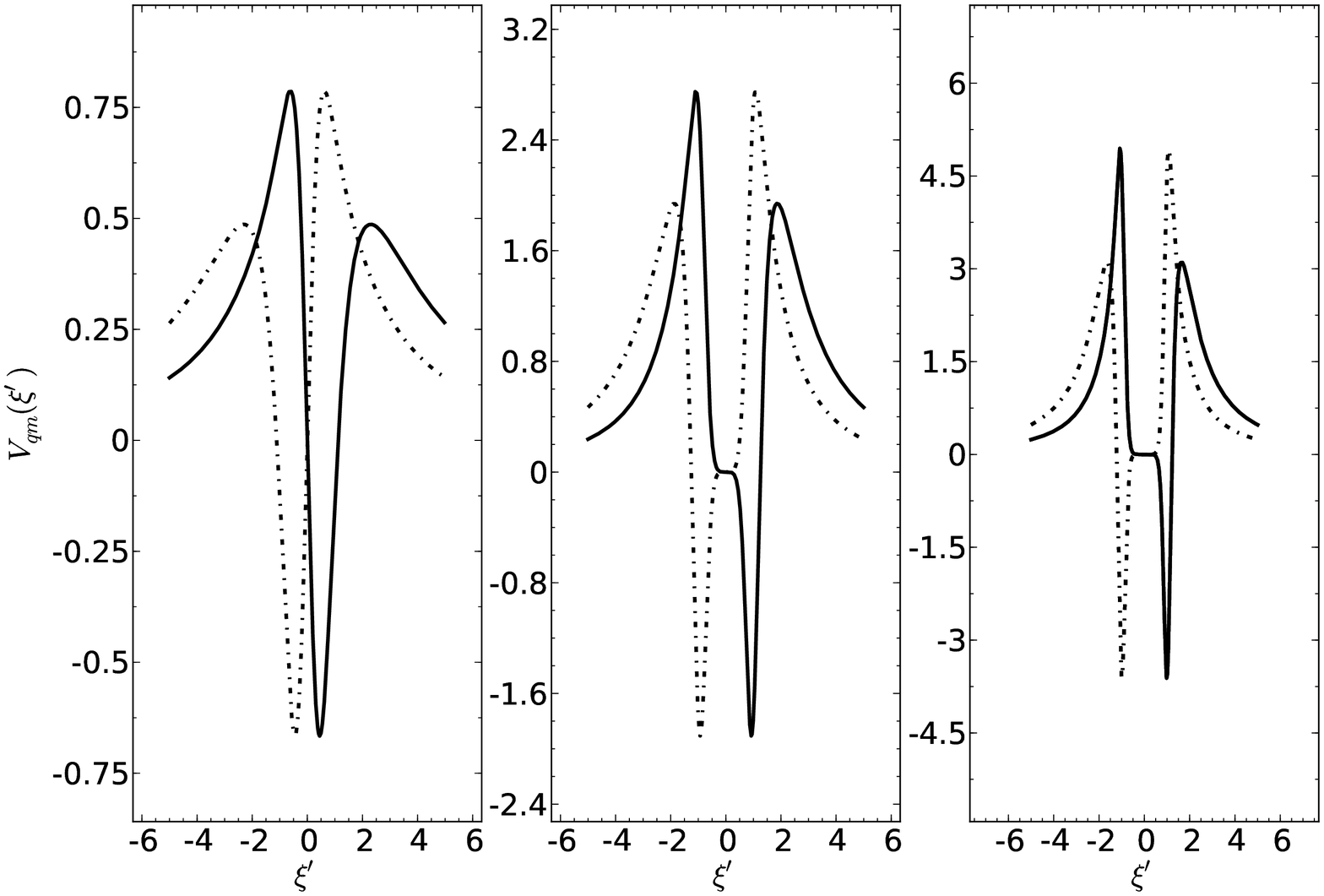}
   \end{center}
   \caption{Quantum Mechanical potential of profiles on a DDW for quadratic scalar-fermion coupling. For $s\in \{1,3,5\}$}
  \label{fig:quad-Yukawa}
\end{figure}

\begin{figure}
   \begin{center}
       \includegraphics[scale=0.33]{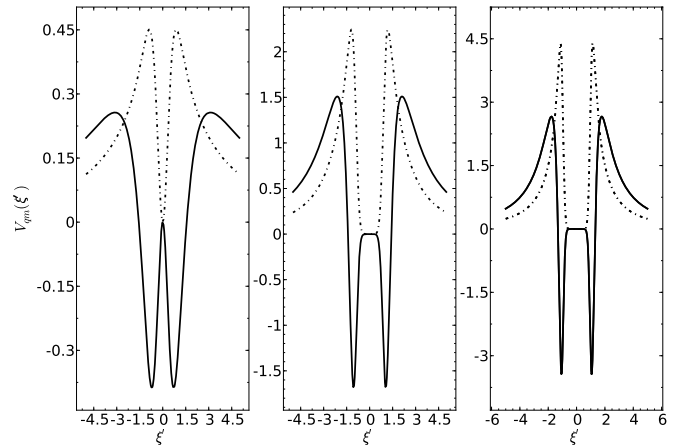}
   \end{center}
   \caption{Quantum Mechanical potential of profiles on a DDW for cubic scalar-fermion coupling. For $s\in \{1,3,5\}$}
   \label{fig:cubic-Yukawa}
\end{figure}

The quantum mechanical potentials, for 5-dimensional massless fermions, and several choices of the polynomial $\mathcal{P}(\phi)$, are shown in figures \ref{fig:linear-Yukawa}, \ref{fig:quad-Yukawa} and \ref{fig:cubic-Yukawa}. In what follows, the parameter $\alpha$ has been set to one and the solid(dashed) line indicates the quantum mechanical potential for negative(positive) chiralities,  as
function of the extra coordinate $\xi$.

\subsection{Massless 5-Dimensional Fermions}

Following the asymptotic behavior of the profiles,
\begin{equation}
  f_\pm \sim e^{\int d\xi \;\Xi(\xi)},
\end{equation}
as proposed in \cite{MPT}, it follows that,
\begin{equation}
\lambda > 2\alpha \left(\frac{2s}{\pi\sqrt{6s-3}}\right)^l,
\end{equation}
with $l$ the minimum (maximum) power of the polynomial $\mathcal{P}(\phi)$ if $s\leq 13$ ($s\geq 15$).

For the plots presented above, $\lambda$ is set to
\begin{equation}
  \lambda = 3\alpha \left(\frac{2s}{\pi\sqrt{6s-3}}\right)^m,
\end{equation}
unless it is  specified otherwise.

\begin{figure}
   \begin{center}
       \includegraphics[scale=0.33]{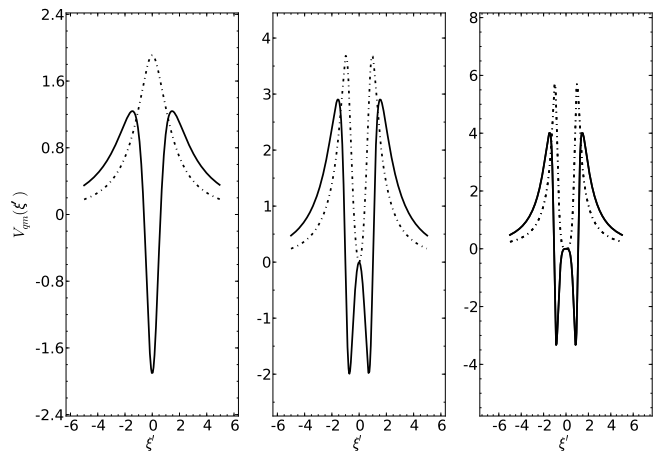}
   \end{center}
   \caption{Quantum Mechanical potential of profiles on a DDW for linear scalar-fermion coupling. For $s\in \{1,3,5\}$}
  \label{fig:linear-Yukawa}
\end{figure}

\begin{figure}
  \begin{center}
    \includegraphics[scale=0.6]{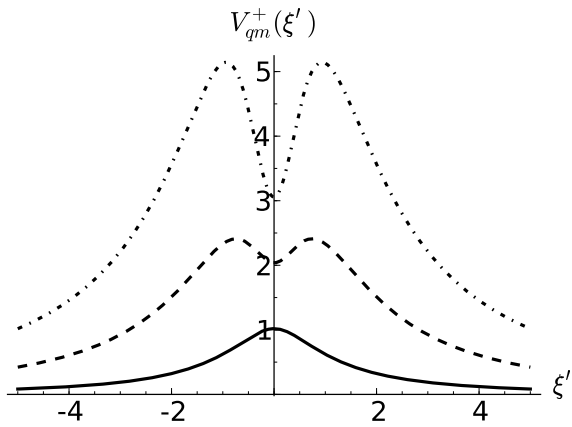}
  \end{center}
  \caption{Quantum Mechanical potential of positive chirality profile on a Double DW for linear scalar-fermion coupling, $s$=1 and $\lambda$ = 1 (solid-), 2 (dashed-) and 3 (dot-dashed line).}
  \label{fig:DDD-Pot+-s=1-l=123}
\end{figure}

For linear scalar-fermion coupling, from figure \ref{fig:linear-Yukawa} it is clear that the positive chirality fermion cannot be localized for $s=1$. However, for higher values of the coupling constant \footnote{The critical value is $\lambda_c=\frac{5\sqrt{3}}{6}$.}, a local well-potential appears around $\xi=0$ (see figure \ref{fig:DDD-Pot+-s=1-l=123}). The well gets steeper as the Yukawa coupling increases.
\begin{table}[ht]

\end{table}%

A very interesting consequence of this is that the  $\sim10^6$ factor of the ratio between the Yukawa couplings of electron and neutrino could explain why right-handed neutrinos are not seen, while right-handed electrons are. The argument is as follows. The  mass difference between electrons and neutrinos, explained by the  Higgs mechanism, is due to the difference of their  Yukawa couplings. Assume that right-handed fermions have the same mass as their left-handed partner. Independently of the Yukawa coupling, left handed fermions are localized because of the well potential $V^-_{qm}(\xi)$. However, $V^+_{qm}(\xi)$ for small Yukawa coupling (neutrino case) is a barrier potential which does not allow localization, whilst for large values of the Yukawa coupling (electron case), the appearance of a local minimum around the brane position favors the localization.

For quadratic scalar-fermion coupling, the quantum mechanical
potential is not $\Z_2$-symmetric, so it is not physically relevant
because the initial problem is clearly symmetric under the exchange
$\xi\to-\xi$. The same argument can be applied to any even power of the
scalar-fermion coupling, so they can be excluded from the analysis.

The analysis of the `spectrum' for generalized odd scalar-fermion
coupling is quite similar to the linear one, except that the
``hill'' potential, which forbids the localization of fermions for
$\lambda<\lambda_c$, is not present.

A more detailed analysis of the spectrum of massless fermions in
these family of domain wall\footnote{And the generalisation of the
family presented in \cite{Guerrero:2005aw,Guerrero:2005xx}}  was
presented  recently in \cite{Liu:2009dw}.

\subsection{Massive 5-Dimensional Fermions}

On this section a numerical analysis of the moduli space is done. Such analysis allows to determine, point-by-point on the moduli, whether a tachyon state exists. Existence of tachyonic states turns into a phenomenological instability of the theory, which permits to rules out such a region of the moduli space, as part of a physical model.

The numerical analysis is done for the action (\ref{Dirac-act}), with $\lambda\ge 1.44$, which assures well-behaved asymptotic limits for  left-handed fermions, as determined in \cite{MPT}. Both chiralities are analyzed independently.


For solving the profile equations, a Numerov algorithm was implemented in Python and Sage \cite{sage}. It  proceeded as follows. First, a lattice was constructed in the moduli space. The considered lattice covers $\lambda\in [1.45, 5.00]$ in steps of $0.05$, and $M\in [0.00,1.00]$ in steps of $0.02$. Secondly, the program finds $V^\pm_{qm}(\xi'=0)$, and from those reference start eigen-values (which represents the 4-dimensional mass square), solves the Schr\"odinger-like equation for the respective ``chiral'' profile. Then, the value of the 4-dimensional mass square is swept by incrementing it in $0.2$ units, until a range of $\Delta m^2\sim 0.2$ is found, where the ground state mass eigen-value lies. Next, the bisection algorithm is applied for improving the precision  of the 4-dimensional mass-square value. Finally, when the 4-dimensional mass-square for the ground state is found with an enhanced precision, one might conclude whether it is a tachyonic state, at that point in the moduli.
\begin{table}
  \begin{center}
    \caption{Resume of the moduli space analysis.}\label{tab:M}
    \begin{tabular}{|>{$}c<{$}|>{$}c<{$}|>{$}c<{$}|>{$}c<{$}|>{$}c<{$}|}
      \hline
      t & s & f^-  &f^+ \\ \hline \hline
      1 &1 & 1.44<\lambda<3.14  & ----         \\ \hline
        &3 & 1.44<\lambda<3.44  & 1.44<\lambda \\ \hline
        &5 & 1.44<\lambda<3.94  & 1.44<\lambda \\ \hline
      3 &1 &      ----          & 1.44<\lambda \\ \hline
        &3 &      ----          & 1.44<\lambda \\ \hline
        &5 & 1.44<\lambda<1.64  & 1.44<\lambda \\ \hline
      5 &1 &      ----          & 1.44<\lambda \\ \hline
        &3 &      ----          & 1.44<\lambda \\ \hline
        &5 &      ----          & 1.44<\lambda \\ \hline
    \end{tabular}
  \end{center}
\end{table}

The scalar-fermion coupling considered below are monomial functions of $\phi$, say, $\mathcal{P}(\phi)=\phi^t$, with $t\in\{1,3,5\}$. Besides, the parameter of the double walls takes similar values.

\subsubsection{Linear Coupling}

When the scalar-fermion coupling is linear ($t=1$), the value of
$\lambda$ should be less than $3.14\pm0.5$ for any value of $M$, in
order for the left-handed ground state to be physically possible.
The right-handed mode does not impose any constrain on $\lambda$.

Changes in the $s$-parameter induce changes in $\lambda_{max}$,
which also varies slowly with $M$. Additionally, for large $s$,
$\lambda_{max}$ increases with $M$.

As was mentioned above, there exists an energy gap between the left- and right-handed ground states. 

Interestingly, in  the simplest case, $s=1$, there is no possibility
of ruling out the tachyonic state for the right-handed chirality.
This fact somehow points to a non-minimalist model for
phenomenology. All other values of $s$ possess a tachyon free
spectrum.

\subsubsection{Cubic Coupling}

In the cubic coupling case, $\mathcal{P}(\phi)=\phi^3$, the
right-handed profiles impose no constrain on the moduli space,
whilst for $s=\{1,3\}$ the tachyonic state is always present.

For $s=5$ there exists a small window in the moduli space which
allows a tachyon free spectrum, given roughly by
$\lambda_{max}\gtrsim 1.64 $.

\subsubsection{Fifth Coupling}

When the fifth coupling was considered, besides the fact that the
right-handed profile imposes no constrain on the moduli space, it
was not possible to get a tachyonic free spectrum for the
left-handed state.

\subsection{Massive 5D Fermions (Revised)}


In odd number of dimensions, it is well known that there is a parity anomaly \cite{Niemi:1983rq,Redlich:1983dv}. In order to avoid this anomaly, the fermion mass in odd dimensions should be multiplied by a sign function of the scalar field \cite{Grossman:1999ra}. Thus, the action is,
\begin{equation}
S = \int\;d^4x\;d\xi\;\sqrt{|g|}   \bar{\Psi}\left(-\fy\nabla -M sgn(\phi)+\lambda\mathcal{P}(\phi)\right)\Psi=0.\label{Dirac-anomaly-free}
\end{equation}

A numerical analysis similar to the one in the preceding subsection is done, but considering the action (\ref{Dirac-anomaly-free}) instead of (\ref{Dirac-act}). 

\subsubsection{Negative Chiralities}

For negative chiralities profiles, once more, the case with $t=s=1$ has  a different behavior than the others. Tachyonic states were found all over the moduli space, except for $M=0$ and any  value of $\lambda$, or for $M=0.02$ if $\lambda\ge 4.65$.

Any other choice of the parameters $M$ and $\lambda$ gives tachyon-free models. It is worth mentioning that choices $(t,s)\in\{(1,3), (1,5), (3,1) \}$ show ground states with $m^2$ close to zero ($m^2\sim 10^{-6}-10^{-8}$), although this value is far from the numerical error ($\Delta m^2\sim 10^{-14}$).

Due to what seems to be numerical problems in the precision, the calculations with $t=5$ could not be done.


\subsubsection{Positive Chiralities}

On the other hand, positive chirality profiles present no tachyonic states, at least  for the considered values of $s$, $t$, $M$ nor $\lambda$. Thus, the ground state has $m^2>0$, but its localization is meta-stable. Besides the fact that meta-stable states rise, the question of whether these are long-lived enough,  has been left aside because the main line of this work rests on formal rather than phenomenological aspects of the theory \footnote{It has been to the authors that in \cite{Ringeval:2001cq}, a similar analysis to the one presented here was done. However, the set up is different and constraints found there were reach by trying to make the model compatible with phenomenology.}.


\section{Discussion and Conclusion}

First of all, a new kind of condition on the 4-dimensional fermions was considered,
\begin{equation}
  \fy\nabla\psi_\pm = 0,
\end{equation}
for finding the equations for the localized profiles, $f_\pm(\xi)$.
Although only the conditions for localizing modes were given, this might be important when multi-brane worlds are considered, as well as non-flat branes. 

The developing of a relationship between the AdS/CFT correspondence and brane-worlds (see \cite{Soda:2010si} and references there in) could add importance to this particular sort of conditions, since the correspondence applies for a stack of $N$ $D$-branes. Moreover, in this scenario, the symmetry breaking might be interpreted geometrically as parallel branes slightly separated in the extra-dimension. This last set-up is exactly one kind of scenario where the new conditions might be important. 

Generalizations of the profile equations were found, by allowing the
scalar-fermion coupling to be non-linear, say a polynomial
$\mathcal{P}(\phi)\bar{\Psi}\Psi$. When $\mathcal{P}(\phi)$ is a
monomial, the physically relevant powers are odd \footnote{In the
single brane solution.}, because for  even powers on $\phi$ the
quantum mechanical potential is not symmetric around the hyperplane
$\xi=0$.

Polynomial couplings get complicated as soon as binomials are
considered, due to the scalar coefficients which regulate the form
of the quantum mechanical potential. Focused  on the double domain
wall family, where $s\in 2\mathbb{Z} +1$ labels the members, the
condition for localizing one of the chiralities is
\begin{equation}
  \lambda > 2\alpha\left(\frac{2s}{\pi\sqrt{6s-3}}\right)^l.
\end{equation}
with $l$ the minimum (maximum) power of the polynomial $\mathcal{P}(\phi)$ if $s\leq 13$ ($s\geq 15$).

When binomial scalar-fermion coupling are considered,
\begin{equation}
   \mathcal{P}(\phi)=c_1\phi^{t_<}+c_2 \phi^{t_>},
\end{equation}
it is straightforward to note that both $t_<$ and $t_>$ should be
even if the symmetric single brane scenario is examined. However,
there is a chance that odd powers of $\phi$ could be
phenomenologically relevant for asymmetric or multi-brane solutions.

Some constraints have been obtain for localizing both chiralities of 4-dimensional fermions coming from massive or  massless 5-dimensional ones. It was found that the simplest model, $s=t=1$, does not accept localization of right chiralities, if the action (\ref{Dirac-act}) is consider. However, when analyzing the action (\ref{Dirac-anomaly-free}), both chiralities could possibly be localized, and it would provide a natural explanation of the  difference between the left- and right-handed neutrino mass, in those models beyond standard one where right-handed neutrinos are likely to exist, as well as the mass difference between different families of fermions. For non-linear scalar-fermion the tachyon-less constrain is made more restrictive as the non-linearity increases.

When a 5-dimensional massive fermion is considered, a slight
correlation between $M$ and $\lambda$ was found, in the studied
region ($0.00<M<1.00$ and $1.45<\lambda<5.00$). The considered region might
be extended, but covering a larger moduli space goes beyond the
scope of this work.

{One might wonder that even when no additional gauge structure is
given, left- and right- handed chiralities behave differently. An
important question to be answered is whether charged particles could
have similar behavior for both chiralities whilst neutrals do not,
because  particles other than neutrinos have the same mass for both
chiralities. In order to follow this line a gauge field, say Abelian
one, must be taken into account. This proposal could be part of
future  research.}

Similar analysis to the one done in this work might be done for the
asymmetric $D$-branes in \cite{MPS,Guerrero:2005aw,Guerrero:2005xx},
some advances towards this are already being done by the authors.

In \cite{Gibbons:2006ge}, the behavior of fermions on (anti-)kink solutions under discrete symmetries is studied. These could be used to complete a theoretical analysis on fermion localization.

\subsection*{Acknowledgement}

We would like to thank A. Melfo, C. Dib and N. Neill for helpful discussions and comments. O.C-F. thanks ICTP-Trieste and CECS-Valvidia for there hospitality during part of the  development of this work.

This work was supported in part by DGIP-UTFSM (Chile), MECESUP$_2$
(Chile) under the grant FSM0605-D3008, and Fondecyt project
11000287.



\appendix

\section{Conventions and Notations}

Through the manuscript, the metrics have signature mostly positive. Since the tetrads formalism is used extensively, distinction between flat and curved coordinates is made  by Latin and Greek indices. Moreover, hated indices run over the whole spacetime whilst unhated ones run over a hypersurface restriction, i.e., on the coordinates parallel to the topological defect.

The gamma matrices are defined in the tangent space, and they satisfy the Clifford algebra,
\begin{equation}
  \left\{\GAM{a},\GAM{b}\right\}=2\eta^{\hat{a}\hat{b}}\mathds{1}.\label{Cliff-alg}
\end{equation}
In even dimensions one can define the chirality matrix $\ga^*$, satisfying the properties
\begin{itemize}
\item $\left\{\GAM{a},\ga^* \right\}=0$.
\item $(\ga^*)^2=\mathds{1}$.
\end{itemize}
From this, the projector operators,
\begin{equation}
  P_\pm= \frac{\mathds{1}-\ga^*}{2},
\end{equation}
are both non-trivial.

In any dimension one may define a set of generators of the Lorentz algebra, constructed with the elements of the Clifford algebra (\ref{Cliff-alg}). These generators of the Lorentz algebra are,
\begin{equation}
  \J^{\hat{a}\hat{b}}=-\frac{\imath}{4}\left[\GAM{a},\GAM{b}\right].
\end{equation}

When the Dirac-Feynman slash notation is used it must be interpreted as,
\begin{equation}
  \fy\partial=\VIN{\mu}{a}\GAM{a}\PA{\mu}.
\end{equation}

The covariant derivative for gauge theories is defined by,
\begin{equation}
  \NAB{\mu}=\PA{\mu}-\imath g A_{\hat{\mu}}-\imath \Omega_{\hat{\mu}},
\end{equation}
with $\Omega$ the gravitational connection, which is related to the Christoffel connection for integer spin fields, and with the spin connection for semi-integer spin fields. Clearly the Dirac-Feynman slash notation can be used with the covariant derivative,
\begin{equation}
  \fy\nabla=\VIN{\mu}{a}\GAM{a}\NAB{\mu}=\VIN{\mu}{a}\GAM{a}\left(\PA{\mu}-\imath g A_{\hat{\mu}}-\imath \Omega_{\hat{\mu}} \right).
\end{equation}

\begin{thebibliography}{10}

\bibitem{RS1}
Lisa Randall and Raman Sundrum.
\newblock {A large mass hierarchy from a small extra dimension}.
\newblock {\em Phys. Rev. Lett.}, 83:3370--3373, 1999.

\bibitem{RS2}
Lisa Randall and Raman Sundrum.
\newblock {An alternative to compactification}.
\newblock {\em Phys. Rev. Lett.}, 83:4690--4693, 1999.

\bibitem{Akama:1982jy}
  K.~Akama,
  Lect.\ Notes Phys.\  {\bf 176} (1982) 267
  [arXiv:hep-th/0001113].

\bibitem{RS}
V.~A. Rubakov and M.~E. Shaposhnikov.
\newblock {Do We Live Inside a Domain Wall?}
\newblock {\em Phys. Lett.}, B125:136--138, 1983.

\bibitem{Visser:1985qm}
  M.~Visser,
  Phys.\ Lett.\  B {\bf 159}, 22 (1985)
  [arXiv:hep-th/9910093].

\bibitem{Kaluza}
Theodor Kaluza.
\newblock {On the Problem of Unity in Physics}.
\newblock {\em Sitzungsber. Preuss. Akad. Wiss. Berlin (Math. Phys. )},
  1921:966--972, 1921.

\bibitem{Klein}
O.~Klein.
\newblock {Quantum theory and five-dimensional theory of relativity}.
\newblock {\em Z. Phys.}, 37:895--906, 1926.

\bibitem{PopeKK}
C.~N. Pope.
\newblock Kaluza-klein theory.
\newblock Technical report, TAMU.
\newblock \href{http://faculty.physics.tamu.edu/pope/ihplec.pdf}{Lecture
  Notes}.

\bibitem{Maldacena1}
Juan~Martin Maldacena.
\newblock {The large N limit of superconformal field theories and
  supergravity}.
\newblock {\em Adv. Theor. Math. Phys.}, 2:231--252, 1998.

\bibitem{Maldacena2}
Ofer Aharony, Steven~S. Gubser, Juan~Martin Maldacena, Hirosi Ooguri, and Yaron
  Oz.
\newblock {Large N field theories, string theory and gravity}.
\newblock {\em Phys. Rept.}, 323:183--386, 2000.

\bibitem{Kogan:2000xc}
Ian~I. Kogan, Stavros Mouslopoulos, Antonios Papazoglou, and Graham~G. Ross.
\newblock {Multi-brane worlds and modification of gravity at large scales}.
\newblock {\em Nucl. Phys.}, B595:225--249, 2001.

\bibitem{Gremm:1999pj}
Martin Gremm.
\newblock {Four-dimensional gravity on a thick domain wall}.
\newblock {\em Phys. Lett.}, B478:434--438, 2000.

\bibitem{MPS}
Alejandra Melfo, Nelson Pantoja, and Aureliano Skirzewski.
\newblock {Thick domain wall spacetimes with and without reflection symmetry}.
\newblock {\em Phys. Rev.}, D67:105003, 2003.

\bibitem{Guerrero:2005aw}
Rommel Guerrero, R.~Omar Rodriguez, and Rafael~S. Torrealba.
\newblock {De Sitter and double asymmetric brane worlds}.
\newblock {\em Phys. Rev.}, D72:124012, 2005.

\bibitem{Guerrero:2005xx}
Rommel Guerrero, R.~Ortiz, R.~Omar Rodriguez, and Rafael~S. Torrealba.
\newblock {De Sitter and double irregular domain walls}.
\newblock {\em Gen. Rel. Grav.}, 38:845--855, 2006.

\bibitem{GMP}
Rommel Guerrero, Alejandra Melfo, and Nelson Pantoja.
\newblock {Self-gravitating domain walls and the thin-wall limit}.
\newblock {\em Phys. Rev.}, D65:125010, 2002.


\bibitem{CastilloFelisola:2004eg}
Oscar Castillo-Felisola, Alejandra Melfo, Nelson Pantoja, and Alba Ramirez.
\newblock {Localizing gravity on exotic thick 3-branes}.
\newblock {\em Phys. Rev.}, D70:104029, 2004.

\bibitem{Kehagias:2000au}
A.~Kehagias and K.~Tamvakis.
\newblock {Localized gravitons, gauge bosons and chiral fermions in smooth
  spaces generated by a bounce}.
\newblock {\em Phys. Lett.}, B504:38--46, 2001.

\bibitem{MPT}
Alejandra Melfo, Nelson Pantoja, and Jose~David Tempo.
\newblock {Fermion localization on thick branes}.
\newblock {\em Phys. Rev.}, D73:044033, 2006.

\bibitem{Liu:2009dw}
Yu-Xiao Liu, Chun-E Fu, Li~Zhao, and Yi-Shi Duan.
\newblock {Localization and Mass Spectra of Fermions on Symmetric and
  Asymmetric Thick Branes}.
\newblock {\em Phys. Rev.}, D80:065020, 2009.

\bibitem{Guerrero:2009ac}
Rommel Guerrero, Alejandra Melfo, Nelson Pantoja, and R.~Omar Rodriguez.
\newblock {Gauge field localization on brane worlds}.
\newblock 2009.

\bibitem{Bajc:1999mh}
  B.~Bajc and G.~Gabadadze,
  ``Localization of matter and cosmological constant on a brane in anti de
  Sitter space,''
  Phys.\ Lett.\  B {\bf 474}, 282 (2000)
  [arXiv:hep-th/9912232].

\bibitem{Papadopoulos:1999tw}
  G.~Papadopoulos, J.~G.~Russo and A.~A.~Tseytlin,
  Class.\ Quant.\ Grav.\  {\bf 17}, 1713 (2000)
  [arXiv:hep-th/9911253].

\bibitem{sage}
W.\thinspace{}A. Stein et~al.
\newblock {\em {S}age {M}athematics {S}oftware ({V}ersion 4.4.1)}.
\newblock The Sage Development Team, 2010.
\newblock {\tt http://www.sagemath.org}.

\bibitem{GP}
  Tony Gherghetta and Alex Pomarol.
  {Bulk Fields and Supersymmetry in a Slice of AdS}
  {\em Nucl.Phys. B}586:141-162, 2000.

\bibitem{Vilenkin-book}
A.~Vilenkin and E.~P.~S. Shellard.
\newblock {\em Cosmic strings and other topological defects}.
\newblock CUP, 1994.

\bibitem{Niemi:1983rq}
A.~J.~Niemi and G.~W.~Semenoff,
 {\em Phys.\ Rev.\ Lett.\ } {\bf 51}, 2077 (1983).

\bibitem{Redlich:1983dv}
A.~N. Redlich.
\newblock {Parity Violation and Gauge Noninvariance of the Effective Gauge
  Field Action in Three-Dimensions}.
\newblock {\em Phys. Rev.}, D29:2366--2374, 1984.

\bibitem{Grossman:1999ra}
Yuval Grossman and Matthias Neubert.
\newblock {Neutrino masses and mixings in non-factorizable geometry}.
\newblock {\em Phys. Lett.}, B474:361--371, 2000.

%
\bibitem{Ringeval:2001cq}
  C.~Ringeval, P.~Peter and J.~P.~Uzan,
  Phys.\ Rev.\  D {\bf 65}, 044016 (2002)
  [arXiv:hep-th/0109194].

\bibitem{Gibbons:2006ge}
  G.~Gibbons, K.~i.~Maeda and Y.~i.~Takamizu,
  Phys.\ Lett.\  B {\bf 647}, 1 (2007)
  [arXiv:hep-th/0610286].

\bibitem{Soda:2010si}
Jiro Soda.
\newblock {AdS/CFT on the brane}.
\newblock 2010.

\end{thebibliography}

\end{document}